\documentstyle[12pt]{article}
\newlength{\pubnumber} 

\catcode`\@=11
\@addtoreset{equation}{section}
\def\section{\@startsection{section}{1}{\z@}{3.5ex plus 1ex minus .2ex}
 {2.3ex plus .2ex}{\large\bf}}
\def\subsection{\@startsection{subsection}{2}{\z@}{2.3ex plus .2ex}
 {2.3ex plus .2ex}{\bf}}

\def\b{{\bf b}}
\def\S{{\bf S}}

\def\bone{{\bf 1}}

\begin{document}

\begin{titlepage}
\samepage{
\setcounter{page}{1}
\rightline{UFIFT-HEP-98-5}
\rightline{\tt hep-ph/9801409}
\rightline{January 1998}
\vfill
\begin{center}
 {\Large \bf  Family Universal Anomalous $U(1)$ \\
     in Realistic Superstring Derived Models\\}
\vfill
 {\large Alon E. Faraggi\footnote{
   E-mail address: faraggi@phys.ufl.edu}\\}
\vspace{.12in}
 {\it Institute For Fundamental Theory, \\
  University of Florida, 
  Gainesville, FL 32611 USA\\}
\end{center}
\vfill
\begin{abstract}
  {\rm
An important issue in supersymmetry phenomenology is
the suppression of squarks contributions to Flavor Changing Neutral
Currents (FCNC).
Recently it was noted that in some free
fermionic three generation models the anomalous $U(1)$ is family universal.
It was further shown that if the $D$--term of the $U(1)_A$
is the dominant source of supersymmetry breaking, the squark masses
are indeed approximately degenerate. In this paper
I discuss the properties of the superstring models that give
rise to the flavor universal anomalous $U(1)$.
The root cause for the universal $U(1)_A$ is the cyclic
permutation symmetry, the characteristic property of the $Z_2\times Z_2$
orbifold compactification, realized in the free fermionic models
by the NAHE set of boundary condition basis vectors.
The properties of the three generation models that preserve this
cyclic permutation symmetry in the flavor charges are discussed.
The cyclic permutation symmetry of the $Z_2\times Z_2$ orbifold
compactification is proposed to be the characteristic property,
of phenomenological interest, that distinguishes it from other
classes of superstring compactifications. 
}
\end{abstract}
\vfill
\smallskip}
\end{titlepage}

\setcounter{footnote}{0}

\def\beq{\begin{equation}}
\def\eeq{\end{equation}}
\def\beqn{\begin{eqnarray}}
\def\eeqn{\end{eqnarray}}
\def\AEF{A.E. Faraggi}
\def\NPB#1#2#3{{\it Nucl.\ Phys.}\/ {\bf B#1} (19#2) #3}
\def\PLB#1#2#3{{\it Phys.\ Lett.}\/ {\bf B#1} (19#2) #3}
\def\PRD#1#2#3{{\it Phys.\ Rev.}\/ {\bf D#1} (19#2) #3}
\def\PRL#1#2#3{{\it Phys.\ Rev.\ Lett.}\/ {\bf #1} (19#2) #3}
\def\PRT#1#2#3{{\it Phys.\ Rep.}\/ {\bf#1} (19#2) #3}
\def\MODA#1#2#3{{\it Mod.\ Phys.\ Lett.}\/ {\bf A#1} (19#2) #3}
\def\IJMP#1#2#3{{\it Int.\ J.\ Mod.\ Phys.}\/ {\bf A#1} (19#2) #3}
\def\nuvc#1#2#3{{\it Nuovo Cimento}\/ {\bf #1A} (#2) #3}
\def\etal{{\it et al,\/}\ }
\hyphenation{su-per-sym-met-ric non-su-per-sym-met-ric}
\hyphenation{space-time-super-sym-met-ric}
\hyphenation{mod-u-lar mod-u-lar--in-var-i-ant}


\setcounter{footnote}{0}

The flavor problem in supersymmetric extensions of the
Standard Model is especially interesting. 
On the one hand the hierarchical pattern of fermion
masses clearly indicates the need for flavor dependent
symmetries. On the other hand the absence of Flavor
Changing Neutral Currents at an observable rate
suggests the need for flavor independent symmetries,
which force squark mass degeneracy. In supersymmetric
field theories the flavor parameters can
be chosen to agree with the data. However, in theories
that aim at the consistent unification of gravity with the gauge interactions
the flavor structure is imposed and cannot be chosen arbitrarily.
Superstring theories are examples of such theories and indeed
it is in general expected that the squark masses in this context
are flavor dependent \cite{il}. The question then arises,
how can there exist the flavor dependent symmetries, needed to
explain the hierarchical fermion mass pattern, while forcing
the degeneracy of squark masses, needed to explain the suppression
of FCNC\footnote{An alternative proposal to the squark mass degeneracy is
the alignment mechanism of Ref. \cite{ns}. A question
of interest is whether such alignment can naturally
arise in a concrete string model}.

In a recent paper the issue of supersymmetry breaking and
squark degeneracy was studied in the three
generation free fermionic superstring models \cite{fp2}.
The proposed mechanism for supersymmetry
breaking is due to the anomalous $U(1)$ $D$--term together with highly
suppressed mass terms for some relevant fields. The effect of
the mass term is to shift the $D$--term of the anomalous $U(1)$
thereby breaking supersymmetry \cite{fayet}. The interesting feature of some
free fermionic models is the fact that the anomalous $U(1)$ is
family universal. 
It was further shown, for a specific choice of flat directions,
that the $D$--terms of the flavor dependent $U(1)$'s vanish in the
minimum of the vacuum.
In these free fermionic models, provided that the
dominant component of the squark masses comes from the anomalous
$U(1)$ $D$--term, the squark masses will be approximately degenerate.

The purpose of this paper is to study the common properties
of the free fermionic models which give rise
to a flavor universal anomalous $U(1)$.
It is expected that once supersymmetry is broken
mass terms of the order of the TeV scale will be generated,
irrespective of the dominant source of supersymmetry breaking.
Such mass terms will then shift the various $D$--terms,
resulting in non--vanishing $D$--term contributions to 
the squark masses.
Therefore, in general, the presence of such non-vanishing
$D$--terms poses a real danger to the viability of the
string models. The generic presence of flavor 
dependent $U(1)$ symmetries in superstring models
then provides an additional criteria in the
selection of the viable models. The solutions
studied in ref. \cite{fp2} provide the guideline
how string models can on the one hand provide
the required symmetries to explain the fermion
mass spectrum while on the other hand explain
the required squark mass degeneracy. 

It should be noted that a universal anomalous $U(1)$
is by no means a general outcome of string solutions
\cite{revamp,fny,alr,lny,lykken,ibanez}.
On the contrary, in a generic superstring model,
we in general expect that the charges of the
chiral generations under the anomalous $U(1)$, like
other potential sources for supersymmetry breaking,
will be flavor dependent. This is demonstrated, for
example, in studies of supersymmetry breaking by
moduli fields, which are in general expected to result
in flavor dependent soft squark masses \cite{il}. The issue
then of flavor universality of the soft squark masses
serves as an important guide in the selection of string vacua.
Understanding this important issue in a specific
class of string vacua, together with other important
phenomenological issues, like the proton longevity and the qualitative
fermion mass spectrum, then serves as a guide to the properties
that an eventual, fully realistic, superstring vacua, might possess.

It should be further emphasized that even in the
restricted class of three generation free fermionic
models, the emergence of a flavor universal $U(1)$
is by no means the generic situation. Indeed,
of the three generation free fermionic models,
one can find several examples in which the
anomalous $U(1)$ is not family universal
\cite{revamp,fny,alr,lny,lykken}.
The task then is to try to isolate, in the class
of free fermionic three generation models,
the properties of the models that do produce a flavor
independent anomalous $U(1)$. While the flavor sfermion
universality is discussed in this paper only with
respect to the anomalous $U(1)$, it should be
remarked that the generic properties of this
restricted class of models may also result
in flavor universal soft SUSY breaking parameters
in other sectors of the models, that will not be
investigated here, but are worthy of further
investigation.

Let us recall that a model in the
free fermionic formulation is defined by a set of boundary
condition basis vectors, and the associated one--loop
GSO projection coefficients \cite{fff}. 
The massless spectrum is obtained by applying the generalized
GSO projections. A physical state defines a vertex operator
which encodes all the quantum numbers with respect to the 
global and gauge symmetries. Superpotential terms are then
obtained by calculating the correlators between the
vertex operators \cite{cvetic,kln}.

The free fermionic models correspond to orbifold models at 
a fixed point in the Narain moduli space.
The same models can be constructed in the orbifold construction
by specifying the background fields and fixing the radii
of the compactified dimensions at the point which 
corresponds to the free fermionic construction.
This correspondence is important because the free fermionic construction 
facilitates the study of the string vacua, and 
the extraction of the properties of the specific
orbifold vacua that are important from the phenomenological
perspective. The purpose of this paper
is partially to highlight one of these properties. 
Namely, the cyclic permutation symmetry of the
$Z_2\times Z_2$ orbifold compactification,
which is one of the basic reasons for the 
appearance of a flavor universal anomalous $U(1)$
in some free fermionic models. 

The free fermionic models studied
here are constructed in two stages. The first stage consists
of the NAHE set, $\{{\bf1}, S, b_1,b_2,b_3\}$. This 
set of boundary condition basis vectors has
been discussed extensively in the literature \cite{revamp,nahe,slm,classi}.
As the properties of the NAHE set are important
to understand the emergence of a family universal anomalous $U(1)$,
for completeness the main features are shortly emphasized. 
The basis vectors of the NAHE set are defined by
\beqn
 &&\begin{tabular}{c|c|ccc|c|ccc|c}
 ~ & $\psi^\mu$ & $\chi^{12}$ & $\chi^{34}$ & $\chi^{56}$ &
        $\overline{\psi}^{1,...,5} $ &
        $\overline{\eta}^1 $&
        $\overline{\eta}^2 $&
        $\overline{\eta}^3 $&
        $\overline{\phi}^{1,...,8} $ \\
\hline
\hline
      {\bf 1} &  1 & 1&1&1 & 1,...,1 & 1 & 1 & 1 & 1,...,1 \\
          $\S$ &  1 & 1&1&1 & 0,...,0 & 0 & 0 & 0 & 0,...,0 \\
\hline
  $ \b_1$ &  1 & 1&0&0 & 1,...,1 & 1 & 0 & 0 & 0,...,0 \\
  $ \b_2$ &  1 & 0&1&0 & 1,...,1 & 0 & 1 & 0 & 0,...,0 \\
  $ \b_3$ &  1 & 0&0&1 & 1,...,1 & 0 & 0 & 1 & 0,...,0 \\
\end{tabular}
   \nonumber\\
   ~  &&  ~ \nonumber\\
   ~  &&  ~ \nonumber\\
     &&\begin{tabular}{c|cc|cc|cc}
 ~&      $y^{3,...,6}$  &
        $\overline{y}^{3,...,6}$  &
        $y^{1,2},\omega^{5,6}$  &
        $\overline{y}^{1,2},\overline{\omega}^{5,6}$  &
        $\omega^{1,...,4}$  &
        $\overline{\omega}^{1,...,4}$   \\
\hline
\hline
    {\bf 1} & 1,...,1 & 1,...,1 & 1,...,1 & 1,...,1 & 1,...,1 & 1,...,1 \\
    $\S$     & 0,...,0 & 0,...,0 & 0,...,0 & 0,...,0 & 0,...,0 & 0,...,0 \\
\hline
$ \b_1$ & 1,...,1 & 1,...,1 & 0,...,0 & 0,...,0 & 0,...,0 & 0,...,0 \\
$ \b_2$ & 0,...,0 & 0,...,0 & 1,...,1 & 1,...,1 & 0,...,0 & 0,...,0 \\
$ \b_3$ & 0,...,0 & 0,...,0 & 0,...,0 & 0,...,0 & 1,...,1 & 1,...,1 \\
\end{tabular}
\label{nahe}
\eeqn
with `0' indicating Neveu--Schwarz boundary conditions
and `1' indicating Ramond boundary conditions, and
with the following
choice of GSO phases:
\beq
      C\left( \matrix{\b_i\cr \b_j\cr}\right)~=~
      C\left( \matrix{\b_i\cr \S\cr}\right) ~=~
      C\left( \matrix{\bone \cr \bone \cr}\right) ~= ~ -1~.
\label{nahephases}
\eeq
The gauge group after imposing the GSO projections of the NAHE
set basis vectors is $SO(10)\times SO(6)^3\times E_8$.
The three sectors $b_1$, $b_2$ and $b_3$ produce
48 multiplets in the chiral 16 representation of $SO(10)$.
The states from each sector transform under the flavor,
right--moving $SO(6)_j$ gauge symmetries, and under the
left--moving global symmetries. This is evident from
table (\ref{nahe}), as each of the sets of the world--sheet fermions
$\{y^{3,\cdots,6}\vert {\bar y}^{3,\cdots,6},{\bar\eta}^1\}$,
$\{y^{1,2},{\omega}^{5,6}\vert {\bar y}^{1,2},{\bar\omega}^{5,6},
{\bar\eta}^2\}$ and 
$\{{\omega}^{1,\cdots,4}\vert {\bar\omega}^{1\cdots,4},
{\bar\eta}^3\}$, has periodic boundary conditions in each
of the basis vectors $b_1$, $b_2$ and $b_3$, respectively.
Also evident from table (\ref{nahe}) is the cyclic permutation symmetry
between the three sectors $b_1$, $b_2$ and $b_3$, with the
accompanying permutation between the three sets of internal
world--sheet fermions.
This cyclic permutation symmetry is the root cause
for the emergence of flavor universal anomalous $U(1)$
in some free fermionic models. If indeed, as argued in ref. \cite{fp2},
flavor universality of the anomalous $U(1)$ is necessary
for the phenomenological viability of a superstring model,
the NAHE set may turn out to be a necessary component
in a realistic string vacua. This, if correct, is
a remarkable outcome, as it serves to isolate the point
in the moduli space where the true string vacuum may be located.

The NAHE set corresponds to $Z_2\times Z_2$ orbifold compactification. 
This correspondence is demonstrated explicitly by adding to the
NAHE set the boundary condition basis vector $X$,
with periodic boundary conditions for the world--sheet
fermions $\{{\bar\psi}^{1,\cdots,5},{\bar\eta}^1,{\bar\eta}^2,{\bar\eta}^3\}$
and antiperiodic boundary conditions for all others. With
a suitable choice of the generalized GSO projection
coefficients, the $SO(10)$ gauge group is enhanced to $E_6$.
The $SO(6)^3$ symmetries are broken to $SO(4)^3\times U(1)^3$.
One combination of the $U(1)$ symmetries is embedded in $E_6$,
\beq
U(1)_{E_6}={1\over\sqrt{3}}(U_1+U_2+U_3).
\label{u1e6}
\eeq
This $U(1)$ symmetry is flavor
independent, whereas the two orthogonal combinations
\beqn
U(1)_{12} &=& {1\over\sqrt{2}}(U_1-U_2)~~~~;\label{u12}\\
U(1)_\psi &=& {1\over\sqrt{6}}(U_1+U_2-2U_3)\label{upsi}
\eeqn
are flavor dependent. The normalization of the various $U(1)$ combinations
is fixed by the requirement that the conformal dimension of the massless
states still gives ${\bar h}=1$ in the new basis.
The final gauge group in this case is therefore
$E_6\times U(1)^2\times SO(4)^3\times E_8$. The three
sectors $b_j\oplus b_j+X$ $(j=1,2,3)$ now produce 24 multiplets in the
27 representation of $E_6$, which are charged under the flavor
$U(1)$ symmetries Eq. (\ref{u12},\ref{upsi}). The multiplicity of the 
generations arises from the transformation under the flavor left--
and right--moving $SO(4)^3$
symmetries. The same model
is constructed in the orbifold formulation by first
constructing the background metric and antisymmetric tensor
which define the Narain model. Fixing the radii of the six
compactified dimensions at $R_I=\sqrt{2}$, produces
an $N=4$ supersymmetric model with $SO(12)\times E_8\times E_8$
gauge group. Acting with the $Z_2\times Z_2$ twisting on the
compactified dimensions, with the standard embedding, then
produces identical spectrum and symmetries as the free fermionic
model, defined by the set of boundary condition basis vectors
$\{ {\bf1},S, b_1,b_2,b_3,X\}$ \cite{foc}.
The set of complex right--moving
world-sheet fermions $\{{\bar\psi}^{1,\cdots,5},{\bar\eta}^1,{\bar\eta}^2,
{\bar\eta}^3,{\bar\phi}^{1,\cdots,8}\}$,
corresponds to the sixteen dimensional compactified
torus of the heterotic string in ten dimensions, whereas the set of left--
and right--moving real fermions
$\{y,\omega\vert{\bar y},{\bar\omega}\}^{1,\cdots,6}$
corresponds to the six compactified dimensions of the $SO(12)$ lattice,
and $\chi^{1,\cdots,6}$ are their fermionic superpartners. 

In the realistic free fermionic models the $E_6$ symmetry is replaced by 
$SO(10)\times U(1)$. This can be seen to arise in two ways.
The first, which is the one employed traditionally in the literature, is
to substitute the vector $X$ above, with a boundary condition basis
vector (typically denoted as $2\gamma$)
with periodic boundary conditions for the complex
world--sheet fermions $\{{\bar\psi}^{1,\cdots,5},{\bar\eta}^1,
{\bar\eta}^2,{\bar\eta}^3,{\bar\phi}^{1,\cdots,4}\}$. The $E_6$ symmetry
is then never realized explicitly and the right--moving gauge group
in this case is $SO(10)\times U(1)_A\times U(1)^2\times SO(4)^3\times
SO(16)$. The three sectors $b_j$ now produce 24 generations in the
16 representation of $SO(10)$ whereas the sectors $b_j+2\gamma$
now produce 24 multiplets in the vectorial 16 representation of the
hidden $SO(16)$ gauge group. Alternatively, the same model is
generated by starting with the set $\{{\bf1}, S, b_1,b_2,b_3,X\}$.
However, we have a discrete choice in the GSO phase $c(X,\xi)=\pm1$,
where $\xi={\bf1}+b_1+b_2+b_3$.
For one choice we obtain the model with $E_6\times E_8$ 
gauge group. For the second choice, the gauge bosons
from the sector $X$, in the $16\oplus{\overline{16}}$ representation
of $SO(10)$ as well as those from the sector $\xi$ in the 128
representation of $SO(16)$, are projected out from the massless spectrum.
We then obtain the same model as with the set
$\{{\bf1},S,b_1,b_2,b_3,2\gamma\}$. The $E_6\times E_8$ gauge group
in this case is therefore broken to $SO(10)\times U(1)_A\times SO(16)$
where $U(1)_A$ is the anomalous $U(1)$ combination. We therefore
see how in this case the anomalous $U(1)$ is just the combination
which is embedded in $E_6$ and its flavor universality is fact 
arises for this reason. 

We recall however that the NAHE set and the related
$E_6\times E_8$ and $SO(10)\times U(1)_A\times S0(16)$ models
are just the first stage in the construction of the 
three generation free fermionic models. The next step
is the construction of several additional boundary condition basis
vectors. These additional boundary condition basis vectors
reduce the number of generations to three generations,
one from each of the sectors $b_1$, $b_2$ and $b_3$.
The additional boundary condition basis vectors break
the $SO(10)$ gauge group to one of its subgroups and
similarly for the hidden $SO(16)$ gauge group. At the same time
the flavor $SO(4)^3$ symmetries are broken to factors of $U(1)$'s.
The number of these $U(1)$'s depends on the specific
assignment of boundary conditions for the set of internal
world--sheet fermions $\{y,\omega\vert{\bar y},{\bar\omega}\}^{1,\cdots,6}$
and can vary from 0 to 6. The additional right--moving $U(1)$ symmetries
arise by pairing two of the right--moving real internal fermions
from the set $\{{\bar y},{\bar\omega}\}$, to form a
complex fermion. For every right--moving
$U(1)$ symmetry, there is a corresponding left--moving
global $U(1)$ symmetry that is obtained by pairing the corresponding
two left--moving real fermions from the set $\{y,\omega\}$.
Each of the remaining world--sheet left--moving real fermions from the set
$\{y,\omega\}$ is paired with a right--moving real fermion
from the set $\{{\bar y},{\bar\omega}\}$ to form an Ising model
operator. The different combinations of the real world--sheet fermions
into complex fermions, or Ising model operators, are determined
by the boundary conditions assignments in the additional boundary
condition basis vectors.
The allowed pairings are constrained
by the requirement that the left--moving world--sheet super--current
of the $N=2$ algebra transforms appropriately under the assignment
of boundary conditions.
In the models that utilize
only periodic and anti--periodic boundary conditions
for the left--moving sector, the eighteen left--moving fermions
are divided into six triplets in the adjoint representation
of the automorphism group $SU(2)^6$ \cite{fff,dlnr},
typically denoted by $\{\chi_i,y_i,\omega_i\}$
$i=1,\cdots,6$.
The six $\chi^{1,\cdots,6}$ are
paired to form the three complex fermions of the NS/R fermions.
The allowed boundary conditions of each of these six triplets
depend on the boundary condition of the world--sheet fermions
$\psi_{1,2}^\mu$. For sectors with periodic boundary conditions,
$b(\psi^\mu_{1,2})=1$,
{\it i.e.} those that produce space time fermions the allowed
boundary condition in each triplet are $(1,0,0)$, $(0,1,0)$
$(0,0,1)$ and $(1,1,1)$. For sectors with antiperiodic
boundary conditions, $b(\psi^\mu_{1,2})=0$,
${\it i.e.}$ those that produce space--time
bosons, the allowed boundary conditions are $(1,1,0)$, $(1,0,1)$
$(0,1,1)$ and $(0,0,0)$.
The super current constraint and the various desirable
phenomenological criteria then limit the possible
complex or Ising model combinations of the left--moving fermions.

In the type of models that are considered here a pair of real
fermions which are combined to form a complex fermion
or an Ising model operator must have the identical boundary
conditions in all sectors. In practice it is sufficient
to require that a pair of such real fermions
have the same boundary conditions in all the boundary
basis vectors which span a given model. The NAHE set of
boundary condition basis vectors already divides the eighteen
left--moving real fermions into three groups
\beqn
&& \{(\chi^{1},~~~,~~~),(\chi^2,~~~,~~~),(~~~,y^{3},~~~),(~~~,y^4,~~~),
				  (~~~,y^5,~~~),(~~~,y^6,~~~)\}~~~~\\
&& \{(~~~,y^{1},~~~),(~~~,y^2,~~~),(\chi^{3},~~~,~~~),(\chi^{4},~~~,~~~),
			(~~~,~~~,\omega^5),(~~~,~~~,\omega^{6})\}~~~~\\
&& \{(~~~,~~~,\omega^1),(~~~,~~~,\omega^2),(~~~,~~~,\omega^3),
		(~~~,~~~,\omega^4),(\chi^{5},~~~,~~~),(\chi^6,~~~,~~~)\}~~~~
\label{reallmft}
\eeqn
where the notation emphasizes the division of the
eighteen left--moving internal world--sheet fermions
into the $SU(2)^6$ triplets.
The $\chi^{12,34,56}$ are the complexified combinations
which generate the $U(1)$ current of the $N=2$ left--moving
world--sheet supersymmetry \cite{kln}. We have the freedom
to complexify all, some or none of the remaining twelve
left--moving world--sheet fermions.
These different choices will in turn produce superstring
models with substantially different phenomenological
implications. 

The additional boundary condition basis vectors, beyond the NAHE set,
 may in fact destroy the cyclic permutation
symmetry between the sectors $b_1$, $b_2$ and $b_3$.
For example if only one $U(1)$ symmetry remains unbroken
from the $SO(4)^3$ symmetries, as for example is the case
in the flipped $SU(5)$ model of ref. \cite{revamp} and
in one of the standard--like models in ref. \cite{slm},
then evidently the cyclic permutation symmetry, which exist
at the level of the NAHE set, is lost.
Thus, we note the first condition for preservation of the cyclic
permutation  symmetry of the NAHE set. Namely, the assignment of
boundary conditions in the additional basis vectors must be such that
the extra $U(1)'s$ respect the permutation symmetry. Therefore,
it is seen that only models with zero, three or six extra
$U(1)$'s that arise from the $SO(4)^3$ group factors,
can preserve the permutation symmetry. 
Furthermore, the
vector combination of the additional boundary condition basis vectors,
combined with the NAHE set basis vectors, can give rise to additional massless
spectrum that contributes to the total trace of the anomalous $U(1)$
charge. Then in the most general case we in fact may expect that the
permutation symmetry is not maintained and therefore that the
anomalous $U(1)$ is not family universal. Nevertheless, 
for some choices of the boundary conditions in the
additional boundary condition basis vectors the
permutation symmetry in the gauge sector is preserved and
in these cases the anomalous $U(1)$ is family universal.
Our aim is therefore to identify the
choices of additional boundary condition basis vectors
that preserve the permutation symmetry and the flavor
universality of the anomalous $U(1)$. 

The free fermionic models that give rise to a flavor universal
$U(1)$ are the two classes of models, \cite{eu} and \cite{top}
that were investigated in ref. \cite{fp2}. The reason for referring to 
those as classes of models is because in addition to the
choices of boundary condition assignments for the free world--sheet
fermions, we still have the freedom, up to the modular invariance
constraints, of the discrete choices of GSO phases. Therefore, each
choice of boundary condition basis vectors still spans a space
of models that are distinguished by the choices of GSO phases
and, in general, may differ in their massless spectrum.
The boundary conditions in the basis vectors beyond the NAHE set
that define the model of ref. \cite{eu} are shown in Eq. (\ref{m278}).
\beqn
 &\begin{tabular}{c|c|ccc|c|ccc|c}
 ~ & $\psi^\mu$ & $\chi^{12}$ & $\chi^{34}$ & $\chi^{56}$ &
        $\bar{\psi}^{1,...,5} $ &
        $\bar{\eta}^1 $&
        $\bar{\eta}^2 $&
        $\bar{\eta}^3 $&
        $\bar{\phi}^{1,...,8} $\\
\hline
\hline
  ${\alpha}$  &  0 & 0&0&0 & 1~1~1~0~0 & 0 & 0 & 0 & 1~1~1~1~0~0~0~0 \\
  ${\beta}$   &  0 & 0&0&0 & 1~1~1~0~0 & 0 & 0 & 0 & 1~1~1~1~0~0~0~0 \\
  ${\gamma}$  &  0 & 0&0&0 &
		${1\over2}$~${1\over2}$~${1\over2}$~${1\over2}$~${1\over2}$
	      & ${1\over2}$ & ${1\over2}$ & ${1\over2}$ &
                ${1\over2}$~0~1~1~${1\over2}$~${1\over2}$~${1\over2}$~0 \\
\end{tabular}
   \nonumber\\
   ~  &  ~ \nonumber\\
   ~  &  ~ \nonumber\\
     &\begin{tabular}{c|c|c|c}
 ~&   $y^3{y}^6$
      $y^4{\bar y}^4$
      $y^5{\bar y}^5$
      ${\bar y}^3{\bar y}^6$
  &   $y^1{\omega}^5$
      $y^2{\bar y}^2$
      $\omega^6{\bar\omega}^6$
      ${\bar y}^1{\bar\omega}^5$
  &   $\omega^2{\omega}^4$
      $\omega^1{\bar\omega}^1$
      $\omega^3{\bar\omega}^3$
      ${\bar\omega}^2{\bar\omega}^4$ \\
\hline
\hline
$\alpha$ & 1 ~~~ 0 ~~~ 0 ~~~ 0  & 0 ~~~ 0 ~~~ 1 ~~~ 1  & 0 ~~~ 0 ~~~ 1 ~~~ 1
\\
$\beta$  & 0 ~~~ 0 ~~~ 1 ~~~ 1  & 1 ~~~ 0 ~~~ 0 ~~~ 0  & 0 ~~~ 1 ~~~ 0 ~~~ 1
\\
$\gamma$ & 0 ~~~ 1 ~~~ 0 ~~~ 1  & 0 ~~~ 1 ~~~ 0 ~~~ 1  & 1 ~~~ 0 ~~~ 0 ~~~ 0
\\
\end{tabular}
\label{m278}
\eeqn

Interestingly enough there exist also a flipped $SU(5)$ model
in which the permutation symmetry between the sectors
$b_1$, $b_2$ and $b_3$ is preserved and in which the structure
of the anomalous and anomaly free $U(1)$'s is similar to the
one found in the model of ref. \cite{eu}. This is the flipped
$SU(5)$ of ref. \cite{price}, and the basis vectors (beyond the NAHE set)
defining this model are shown in Eq. (\ref{price}).
\beqn
 &\begin{tabular}{c|c|ccc|c|ccc|c}
 ~ & $\psi^\mu$ & $\chi^{12}$ & $\chi^{34}$ & $\chi^{56}$ &
        $\bar{\psi}^{1,...,5} $ &
        $\bar{\eta}^1 $&
        $\bar{\eta}^2 $&
        $\bar{\eta}^3 $&
        $\bar{\phi}^{1,...,8} $\\
\hline
\hline
  ${b_4}$  &  1 & 1&0&0 & 1~1~1~1~1 & 1 & 0 & 0 & 1~1~1~1~0~0~0~0 \\
  ${b_5}$   &  1 & 0&1&0 & 1~1~1~1~1 & 0 & 1 & 0 & 1~1~1~1~0~0~0~0 \\
  ${\gamma}$  &  0 & 0&0&0 &
		${1\over2}$~${1\over2}$~${1\over2}$~${1\over2}$~${1\over2}$
	      & ${1\over2}$ & ${1\over2}$ & ${1\over2}$ &
                1~1~1~1~${1\over2}$~${1\over2}$~0~0 \\
\end{tabular}
   \nonumber\\
   ~  &  ~ \nonumber\\
   ~  &  ~ \nonumber\\
     &\begin{tabular}{c|c|c|c}
 ~&   $y^4{y}^5$
      $y^3{\bar y}^3$
      $y^6{\bar y}^6$
      ${\bar y}^4{\bar y}^5$
  &   $y^1{\omega}^6$
      $y^2{\bar y}^2$
      $\omega^5{\bar\omega}^5$
      ${\bar y}^1{\bar\omega}^6$
  &   $\omega^2{\omega}^3$
      $\omega^1{\bar\omega}^1$
      $\omega^4{\bar\omega}^4$
      ${\bar\omega}^2{\bar\omega}^3$ \\
\hline
\hline
$b_4$ & 0 ~~~ 1 ~~~ 1 ~~~ 0  & 0 ~~~ 0 ~~~ 1 ~~~ 0  & 0 ~~~ 0 ~~~ 1 ~~~ 0
\\
$b_5$  & 0 ~~~ 0 ~~~ 1 ~~~ 0  & 0 ~~~ 1 ~~~ 1 ~~~ 0  & 0 ~~~ 1 ~~~ 0 ~~~ 0
\\
$\gamma$ & 1 ~~~ 0 ~~~ 0 ~~~ 0  & 0 ~~~ 0 ~~~ 1 ~~~ 1  & 0 ~~~ 0 ~~~ 1 ~~~ 1
\\
\end{tabular}
\label{price}
\eeqn

The difference is that this flipped $SU(5)$ superstring model contains
two additional pairs of $10+\overline{10}$ of $SU(5)$,
beyond those that arise from the NAHE set basis vectors ({\it i.e.} from the
sectors $b_4$ and $b_5$).
In the flipped $SU(5)$ case there is an additional freedom
in the identification of the light generations. 
The final charges of the generations under the anomalous $U(1)$
will depend on this identification. Consequently, whether or
not the universality of the anomalous $U(1)$ can be preserved
in flipped $SU(5)$ models depends on a more detailed analysis
of flat directions and the fermion mass spectrum. 
Nevertheless, we can still identify several common
features which give rise to a similar structure of
the anomalous $U(1)$ in these three models. 

Turning to the next common features.
There is a large amount of freedom in the allowed pairings of the
internal world--sheet fermions,
$\{y,\omega\vert{\bar y},{\bar\omega}\}^{1,\cdots,6}$.
Detailed discussion on the classification of
the models by the world--sheet pairings is given in \cite{classi}.
It is however noted that in all three models
that produced the universal $U(1)_A$, the pairing is similar.
Namely the choice of pairings is such that all six triplets
of the $SU(2)^6$ automorphism group are inter-wind. For example,
in the models of ref. \cite{eu,top} the pairing is:
\beqn
&& \{(y^3y^6,y^4{\bar y}^4,y^5{\bar y}^5,{\bar y^3}{\bar y}^6), \nonumber\\
&& (y^1\omega^5,y^2{\bar y}^2,\omega^6{\bar\omega}^6,
			{\bar y}^1{\bar\omega}^5),\nonumber\\
&& (\omega^2\omega^4,\omega^1{\bar\omega}^1,\omega^3{\bar\omega}^3,
				{\bar\omega^2}{\bar\omega}^4)\}.
\label{278pairings}
\eeqn

In this choice of pairings it is noted that the complexified
left--moving fermions mix the six left--moving $SU(2)$ triplets.
A similar feature is observed in the choice of pairings in the flipped
$SU(5)$ model of ref. \cite{price} (see Eq. (\ref{price})).
The choice of pairing in this model can in fact to be identical
to the one used in the models of ref. \cite{eu,top}, with an 
appropriate change of the boundary conditions in the basis
vector $\alpha$, that still produces the same spectrum.  
In contrast we note that all other
choices of pairings that have been employed in the construction of realistic
free fermionic models have not led to a universal $U(1)_A$ (
see refs. \cite{revamp,fny,alr,lny} for several examples of such models).

As we discussed above the NAHE set possesses an inherent cyclic
permutation symmetry which is a manifestation of the $Z_2\times Z_2$
orbifold compactification with the standard embedding. However, to
construct the three generation free fermionic models we have to 
supplement the NAHE set with three additional boundary condition
basis vectors, typically denoted as $\{\alpha,\beta,\gamma\}$.
In order for the anomalous $U(1)$ to remain universal the
additional boundary conditions $\{\alpha,\beta,\gamma\}$
must then preserve the permutation symmetry of the NAHE set,
in the charges of the chiral generations under the flavor
symmetries. 

The standard--like model of ref. \cite{eu} and the flipped
$SU(5)$ model of ref. \cite{price} exhibit a similar structure
of the anomalous $U(1)$ and anomaly free combinations.
In these two models the $U(1)$ symmetries, generated by the world--sheet
complex fermions $\{{\bar\eta}^1,{\bar\eta}^2,{\bar\eta}^3\}$ and
$\{{\bar y}^3{\bar y}^6,{\bar y}^1{\bar\omega}^5,{\bar\omega}^2
{\bar\omega}^4\}$ (or $\{{\bar y}^4{\bar y}^5,{\bar y}^1{\bar\omega}^6,
{\bar\omega}^2{\bar\omega}^3\}$ in the case of the flipped $SU(5)$
model of ref. \cite{price}) are anomalous, with:
${\rm Tr} U_1=
{\rm Tr} U_2={\rm Tr} U_3=24,{\rm Tr} U_4= {\rm Tr} U_5=
{\rm Tr} U_6=-12$. 
The anomalous $U(1)$ combination in both models is therefore given by
\beq
U_A={1\over{\sqrt{15}}}(2 (U_1+U_2+U_3) - (U_4+U_5+U_6))~;~ {\rm Tr} Q_A=
{1\over{\sqrt{15}}}180~.
\label{u1a}
\eeq
One choice for the five anomaly--free combinations is
given by
\beqn
{U}_{12}&=& {1\over\sqrt{2}}(U_1-U_2){\hskip .5cm},{\hskip .5cm}
{U}_{\psi}={1\over\sqrt{6}}(U_1+U_2-2U_3),\label{u12upsi}\\
{U}_{45}&=&{1\over\sqrt{2}}(U_4-U_5){\hskip .5cm},{\hskip .5cm}
{U}_\zeta ={1\over\sqrt{6}}(U_4+U_5-2U_6),\label{u45uzeta}\\
{U}_\chi &=& {1\over{\sqrt{15}}}(U_1+U_2+U_3+2U_4+2U_5+2U_6).
\label{uchi}
\eeqn
The anomalous $U(1)$,
containing the {\it sums} of $U_{1,2,3}$
and $U_{4,5,6}$
is universal with respect to the
three families from the sectors $b_1$, $b_2$ and $b_3$.
This flavor universality of the anomalous $U(1)$
is thus a consequence of the family permutation symmetry of
the six $U(1)$--interactions. In the model of ref. \cite{top}
only the three $U(1)$ generated by the world--sheet 
complex fermions $\{{\bar\eta}^1,{\bar\eta}^2,{\bar\eta}^3\}$
are anomalous with ${\rm Tr} U_1=
{\rm Tr} U_2={\rm Tr} U_3=24$. The anomalous $U(1)$ combination
in this model is just the combination given in Eq. (\ref{u1e6})
and the two orthogonal anomaly free combinations are those
given in Eqs. (\ref{u12}) and (\ref{upsi}).

The next common feature in all three models that yielded a
universal anomalous $U(1)$ is in the structure of the 
assignment of boundary conditions to the internal world--sheet
fermions $\{y,\omega\vert{\bar y},{\bar\omega}\}^{1,\cdots,6}$.
This common structure is exhibited by the fact that
all three models produce a sector, which is a combination of
the basis vectors $\{\alpha,\beta,\gamma\}$, and possibly with
the NAHE set basis vectors, that can produce, depending on the
choice of generalized GSO projection coefficients, additional space--time
vector bosons. In this vector combination the left--moving sector
is completely Neveu--Schwarz. In the model of ref. \cite{eu} all
the additional space--time vector bosons are projected by the 
generalized GSO projections, while in the model of ref. \cite{top}
the $SU(3)_C$ gauge group is enhanced to a $SU(4)_C$ gauge group.
Similar phenomena is encountered in the model of ref. \cite{price}.
Thus, we see that the additional boundary condition basis vectors in
all three models that resulted in a universal $U(1)_A$ are constructed
near an enhanced symmetry point in the moduli space.
This type of enhanced symmetries may in fact play an
important role in explaining the suppression of proton decay
from dimension four and five operators \cite{pd}.

To conclude, in this paper identified the common
features that yielded the appearance of a family universal $U(1)_A$
in the free fermionic models. These common features
are the NAHE set of boundary condition basis vectors, the
choice of pairing of the real world--sheet fermions
$\{y,\omega\vert{\bar y},{\bar\omega}\}^{1,\cdots,6}$,
and the assignment of boundary conditions in the basis
vectors beyond the NAHE set,to these internal real fermions,
which indicates that these models are near an enhanced symmetry
point.
We further note that family universality of the anomalous $U(1)$
has also been found to be desirable in recent attempts to
fit the fermion mass spectrum by the use of Abelian horizontal
symmetries \cite{ramond}.
In this regard we remark that the realistic
free fermionic model, and their relation to the
underlying $Z_2\times Z_2$ orbifold compactification,
possess a phenomenologically appealing structure, 
which is not shared by other classes of three generation
orbifold models. This include: (a) the natural
emergence of three generations {\it i.e.}
each generation is obtained from one of the
twisted sectors of the $Z_2\times Z_2$ orbifold.
The existence of three generations together,
with the flavor symmetries needed to explain
the fermion mass spectrum are correlated
with the properties of the underlying orbifold
compactification. (b) The standard $SO(10)$
embedding of the weak--hypercharge. This is
an important point that cannot be overemphasized. Although from
the point of view of ordinary GUTs it seems rather trivial,
this is not the case in string theory. Indeed
there exist numerous examples of three
generations string models, in which the 
weak-hypercharge does not have the standard
$SO(10)$ embedding. Standard $SO(10)$ embedding of the weak-hypercharge
means that free fermionic models, despite the fact that a GUT symmetry
does not exist in the effective low energy field theory, still predict
the canonical value for the weak mixing angle, $\sin^2\theta_W=3/8$.
The fact that free fermionic
models yield the standard $SO(10)$ embedding
for the weak--hypercharge enables the agreement 
of these models with the measured values of 
$\alpha_s(M_Z)$ and $\sin_\theta^2(M_Z)$. 
(c) flavor universal anomalous $U(1)$. This
last feature is again a special feature
of free fermionic models that arises due to
the permutation symmetry of the underlying
$Z_2\times Z_2$ orbifold. Furthermore,
as these properties are related to the gauge and
local discrete symmetries of these models they
are expected to survive also in the possible
nonperturbative extension of these models.
If the NAHE set turns out to be
a necessary ingredient for obtaining a universal $U(1)_A$,
and if indeed a universal anomalous $U(1)$ is necessary
to obtain agreement with the phenomenological constraints,
it may be an additional indication, that
the true string vacuum is in the vicinity of the 
$Z_2\times Z_2$ orbifold, at the free fermionic point in the
Narain moduli space, and with the standard embedding of the
gauge connection. 

I thank Jogesh Pati and Pierre Ramond for valuable discussions.
This work was supported in part by 
DOE Grant No.\ DE-FG-0586ER40272.

\bibliographystyle{unsrt}

\vfill
\eject

\end{document}